\documentclass[aps,prl,twocolumn,superscriptaddress]{revtex4-1}
\usepackage{graphicx}
\usepackage{bm}
\usepackage{color}
\usepackage[latin1]{inputenc} 

\newlength\figurewidth
\newcommand\rien[1]{}

\newcommand{\correc}[1]{}

\begin{document}

\title{Optical nonlinearity for few-photon pulses on a quantum dot-pillar cavity device}

\author{V. Loo}
\affiliation{Laboratoire de Photonique et Nanostructures, LPN/CNRS, Route de Nozay, 91460 Marcoussis, France}
\affiliation{Université Paris Diderot, Département de Physique, 4 rue Elsa Morante, 75013 Paris, France}

\author{C. Arnold}
\affiliation{Laboratoire de Photonique et Nanostructures, LPN/CNRS, Route de Nozay, 91460 Marcoussis, France}

\author{O. Gazzano}
\affiliation{Laboratoire de Photonique et Nanostructures, LPN/CNRS, Route de Nozay, 91460 Marcoussis, France}

\author{A. Lemaître}
\affiliation{Laboratoire de Photonique et Nanostructures, LPN/CNRS, Route de Nozay, 91460 Marcoussis, France}

\author{I. Sagnes}
\affiliation{Laboratoire de Photonique et Nanostructures, LPN/CNRS, Route de Nozay, 91460 Marcoussis, France}

\author{O. Krebs}
\affiliation{Laboratoire de Photonique et Nanostructures, LPN/CNRS, Route de Nozay, 91460 Marcoussis, France}

\author{P. Voisin}
\affiliation{Laboratoire de Photonique et Nanostructures, LPN/CNRS, Route de Nozay, 91460 Marcoussis, France}

\author{P. Senellart}
\affiliation{Laboratoire de Photonique et Nanostructures, LPN/CNRS, Route de Nozay, 91460 Marcoussis, France}

\author{L. Lanco}
\email[]{loic.lanco@lpn.cnrs.fr}
\affiliation{Laboratoire de Photonique et Nanostructures, LPN/CNRS, Route de Nozay, 91460 Marcoussis, France}
\affiliation{Université Paris Diderot, Département de Physique, 4 rue Elsa Morante, 75013 Paris, France}

\date{\today}

\begin{abstract}

Giant optical nonlinearity is observed under both continuous-wave and pulsed excitation in a deterministically-coupled quantum dot-micropillar system, in a pronounced strong-coupling regime. Using absolute reflectivity measurements we determine the critical intracavity photon number as well as the input and output coupling efficiencies of the device. Thanks to a near-unity input-coupling efficiency, we demonstrate a record nonlinearity threshold of only 8 incident photons per pulse. The output-coupling efficiency is found to strongly influence this nonlinearity threshold. We show how the fundamental limit of single-photon nonlinearity can be attained in realistic devices, which would provide an effective interaction between two coincident single photons.

\end{abstract}

\maketitle

A single quantum emitter is an extremely non-linear optical medium, where the interaction with a first single photon modifies the transmission probability for a second photon \cite{Birnbaum2005}. Along with various applications, such as the generation of non-classical light \cite{Faraon2008} or the non-destructive measurement of photon number \cite{Imoto1985}, this fundamental property has led to the proposal of single photon switches \cite{Chang2007,Hwang2009}: a gate single photon switches off the single emitter optical response, and prevents or allows the transmission of a signal photon. The main difficulty to implement a single photon switch is to ensure that the gate single photon will interact with the quantum emitter with a close to unity probability. To ensure this optimal coupling, one can couple the quantum emitter to an optical microcavity mode \cite{Auffeves-Garnier2007}, provided that this mode allows an optimal external matching with the incident photons. When the emitter is in the strong-coupling regime with the cavity mode, this also allows fast switching times to be obtained, governed by the photon lifetime in the cavity.

Semiconductor quantum dots (QDs) are promising systems to implement a single photon optical switch in a solid state system. Recently, resonant spectroscopy on coupled QD-cavity devices, in the form of photonic crystals \cite{Englund2007,Reinhard2011,Englund2012,Sridharan2012,Volz2012} or microdisks \cite{Srinivasan2007,Srinivasan2008}, has demonstrated saturation of the QD-emission and fast optical switches. These works all concluded that optical nonlinearity is obtained when close to unity photon numbers are reached \emph{inside} the cavity. However, hundreds of incident photons were required to obtain a single intracavity photon. Distinguishing between the intracavity photon number and the number of incident photons per pulse is crucial for future quantum applications. Indeed, providing a nonlinearity at the level of single \emph{incident} photons would open the way for a large number of realizations in the field of quantum information \cite{Kimble2008}. The effective interaction between two coincident photons could be used for direct single-photon routing \cite{Dayan2008,Hoi2011} or for more sophisticated protocols intended to manipulate the quantum phase \cite{Turchette1995} or the polarization state \cite{Duan2004} of single-photon pulses. In this respect QD-micropillar systems \cite{Rakher2009,Loo2010,Young2011,Arnold2012} are very promising candidates as they allow an externally mode-matched operation with high input and output coupling efficiencies. 

In this Letter we report on the observation of optical nonlinearity in a QD-micropillar system with few incident photons per optical pulse. Continuous-wave (CW) and pulsed reflectivity measurements are performed and allow us to extract all the coupling parameters. A near-unity input-coupling efficiency is obtained. We characterize both the internal and external photon numbers and demonstrate, using short-pulse excitation, a record nonlinearity threshold at 8 incident photons per pulse. Combined experimental and theoretical studies show that the output-coupling efficiency is a crucial parameter to lower the nonlinearity threshold. We anticipate that a nonlinear interaction between two coincident single-photon pulses can be obtained for realistic devices with a near-unity output-coupling efficiency.

The sample under study has been grown by molecular beam epitaxy on a GaAs substrate. Two distributed Bragg reflectors, consisting in alternating $\lambda / 4$ layers of Ga$_{0.9}$Al$_{0.1}$As and Ga$_{0.05}$Al$_{0.95}$As, surround a GaAs $\lambda$ cavity. The bottom and top Bragg mirrors are constituted by 36 and 32 pairs. The GaAs cavity embeds a layer of self-assembled InGaAs QDs, and the number of quantum dots emitting at the planar cavity mode energy has been strongly reduced by shifting the dot energy distribution using rapid thermal annealing. The lateral confinement is obtained by inductively coupled plasma etching, leading to a micropillar with a diameter of $2.1 \mathrm{\mu m}$. Spatial and spectral matching between the cavity mode and a selected QD transition have been achieved thanks to the \emph{in-situ} lithography technique \cite{Dousse2008}: the resulting device is in the strong-coupling regime, as shown using photoluminescence measurements on the same sample in Ref. \cite{Dousse2009}. 

\begin{figure}[h!]
\centering \includegraphics[width=0.49\textwidth]{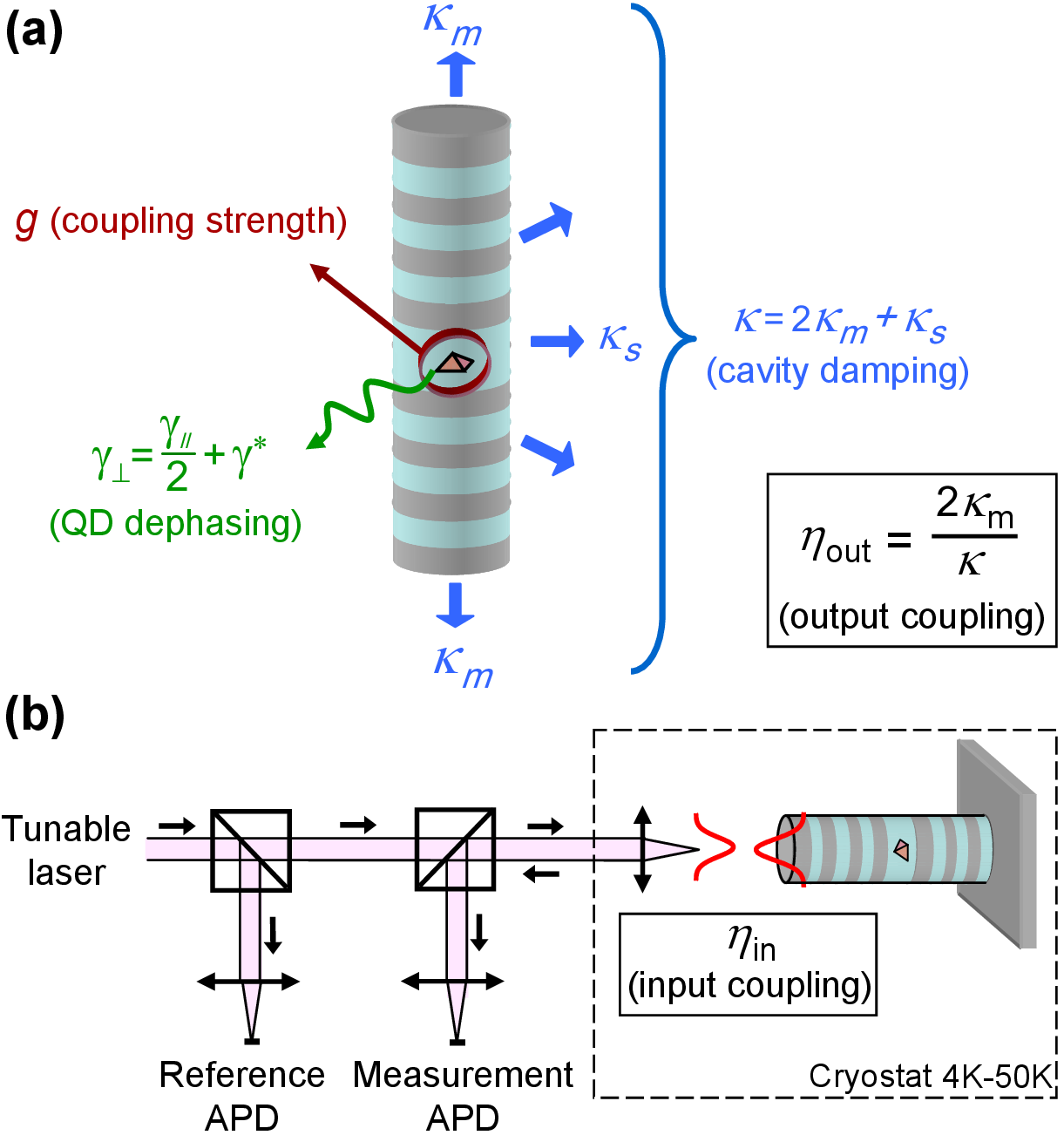}
\caption{\textbf{(a)} Diagram of the QD-cavity device and corresponding physical quantities. \textbf{(b)} Experimental setup. \label{Fig1}}
\end{figure}

Fig.~\ref{Fig1}(a) presents a scheme of the QD-micropillar system and of the physical parameters governing the device properties: $g$ is the QD-mode coupling strength related to the vacuum Rabi splitting, $\kappa$ the intensity damping rate of the cavity, and $\gamma_\bot$ the QD dephasing rate. The top and bottom Bragg mirrors have the same reflectivity and thus the same damping rate $\kappa_m$. The total cavity damping rate is given by the sum $\kappa=2\kappa_m+\kappa_s$, with $\kappa_s$ the side-leakage damping rate induced by the micropillar sidewall roughness. The output coupling efficiency $\eta_\mathrm{out}$ is defined as the probability that a photon in the mode will leave the cavity by one of the desired channels (top or bottom mirror), so that $\eta_\mathrm{out}=\frac{2\kappa_m}{\kappa}$. 

A complete description of the system requires the distinction between the QD dephasing rate $\gamma_\bot$ and the incoherent exciton decay rate $\gamma_{||}$, $\gamma_{\bot}$ and $\gamma_{||}$ being standard notations in atom cavity QED \cite{Armen2006}. The incoherent decay rate $\gamma_{||}$ corresponds to either exciton nonradiative decay or spontaneous emission into other modes than the cavity fundamental mode, both effects leading to a population decay with a loss of information in the environment. This incoherent decay rate $\gamma_{||}$ gives the lifetime-limited contribution to the QD dephasing rate, through $\gamma_\bot=\frac{\gamma_{||}}{2}+\gamma^*$.  The second contribution $\gamma^*$ corresponds to pure dephasing processes affecting the exciton coherence without changing its lifetime. With these notations the dynamics of the QD-cavity system is governed by two dimensionless quantities which are well-known in atom cavity QED \cite{Armen2006,SupplementalMaterials}:
\begin{equation}
 n_c=\frac{\gamma_{||} \gamma_{\bot}}{4g^2} \ \ \ \ \ \mathrm{and} \ \ \ \ \ C=\frac{g^2}{\kappa \gamma_{\bot}}
\end{equation}
The first one is the critical intracavity photon number, i.e. the intracavity photon number at which the nonlinear behaviour becomes significant. The second one is the cooperativity parameter which measures the amplitude of the quantum effects that a single two-level system induces inside an optical cavity \cite{Armen2006}.

The experimental setup is sketched in Fig. ~\ref{Fig1}(b). The sample is placed inside a helium vapor cryostat, altogether with a focusing lens. A CW or pulsed laser is injected into and reflected from the micropillar with a finely tunable photon energy $\hbar \omega$ ($\hbar=1$ units are used in the following). The incident power is measured with a first avalanche photodiode, a second one being used to measure the reflected power. A wavelength meter monitors the excitation energy, making up for any drift or mode-hop of the laser. The input-coupling efficiency $\eta_\mathrm{in}$ is defined as the probability that an incoming photon focused on the micropillar will be coupled to the cavity mode. A near-unity input-coupling efficiency is expected thanks to the very good overlap between the mode and the incident optical beam, as was demonstrated in previous works using the same setup \cite{Loo2010,Arnold2012}. 

Every reflection-transmission coefficient has been characterized, which allows us to deduce within a $5\%$ accuracy the absolute values of the incident and reflected powers, and to distinguish the contribution of interest (power reflected by the micropillar due to light coupled into the optical mode) from the residual contribution due to uncoupled reflected light \cite{SupplementalMaterials}. The effects of the input and output coupling efficiencies can be separated, which in turn allows the deduction of the intracavity photon number from the measured optical powers.

\begin{figure}[ht!]
\centering \includegraphics[width=0.49\textwidth]{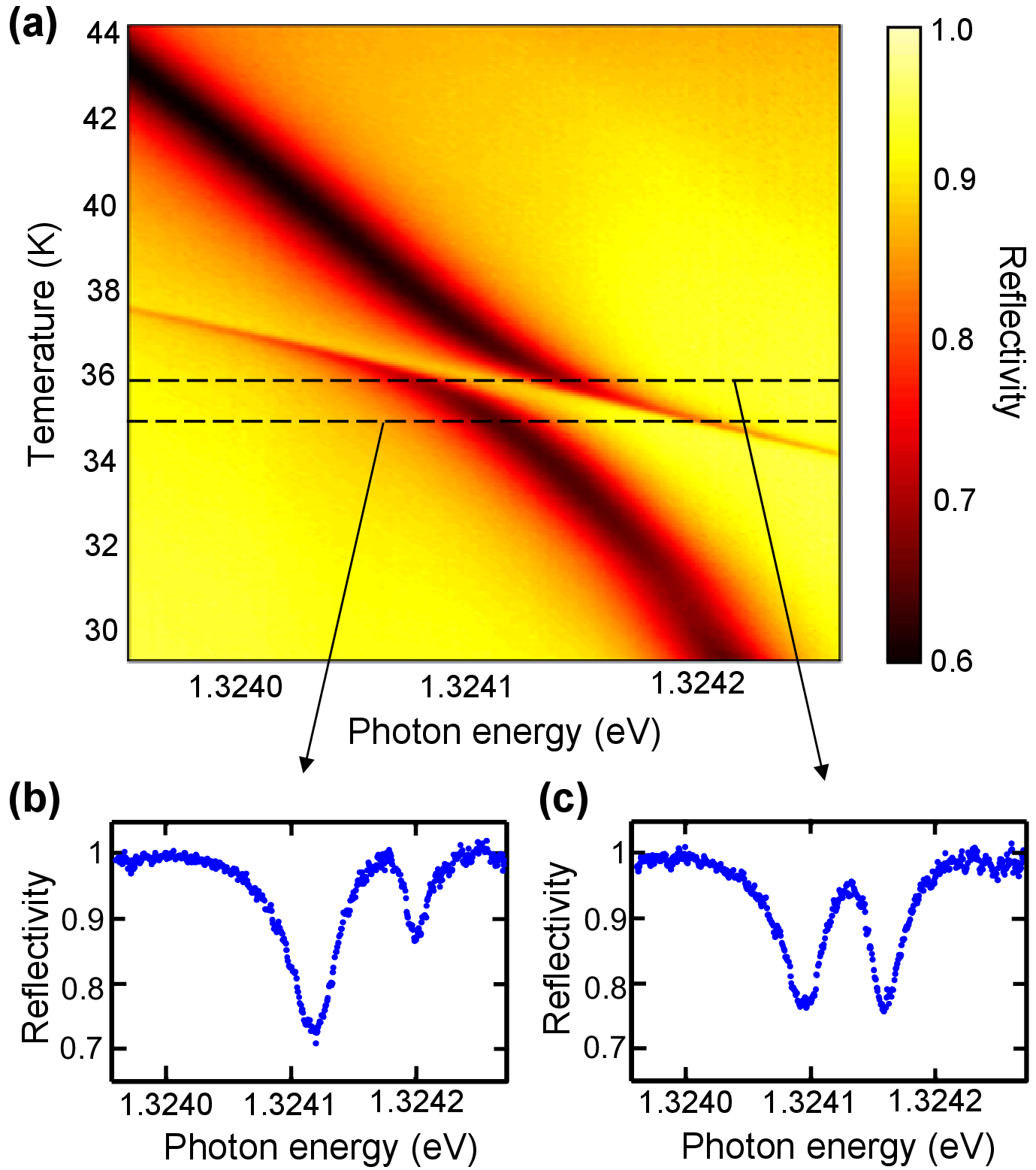}
\caption{\textbf{(a)} Low power reflectivity measurement as a function of photon energy and temperature. \textbf{(b)} Reflectivity spectrum at $T=34.8K$. \textbf{(c)} Reflectivity spectrum at $T=35.9K$. The reflectivity spectra are normalized to unity far from resonance. \label{Fig2}}
\end{figure}

When the sample temperature is increased, the bare QD energy $\omega_{QD}$ and the bare cavity mode energy $\omega_c$ shift with different temperature dependences: this property is used to continuously vary the QD-cavity detuning \cite{Loo2010}. Figure~\ref{Fig2}(a) shows an experimental map of the reflectivity measured as a function of both temperature and photon energy $\omega$, where the darker areas correspond to lower reflectivities. A low incident power $P_0=1.1$~nW has been used to avoid the saturation of the QD transition. The system is in a pronounced strong-coupling regime: in such a case the low-reflectivity dips directly evidence the temperature dependence of the two coupled exciton-photon eigenstates, and their anticrossing when the device temperature is tuned \cite{Loo2010}. Figure~\ref{Fig2}(b) presents a reflectivity spectrum measured at a fixed temperature $T=34.8K$ where the asymmetrical shape arises from the unequal excitonic and photonic parts for the exciton-photon eigenstates. Figure~\ref{Fig2}(c) presents the reflectivity spectrum measured at the resonance temperature $T=35.9K$, where the exciton-photon eigenstates have equal excitonic and photonic parts. When the device is tuned across the resonance the lower-energy eigenstate (resp. higher-energy eigenstate) evolves from cavity-like to QD-like (resp. QD-like to cavity-like) behaviour, in agreement with the Jaynes-Cummings model of cavity QED.

\begin{figure}[ht!]
\centering \includegraphics[width=0.49\textwidth]{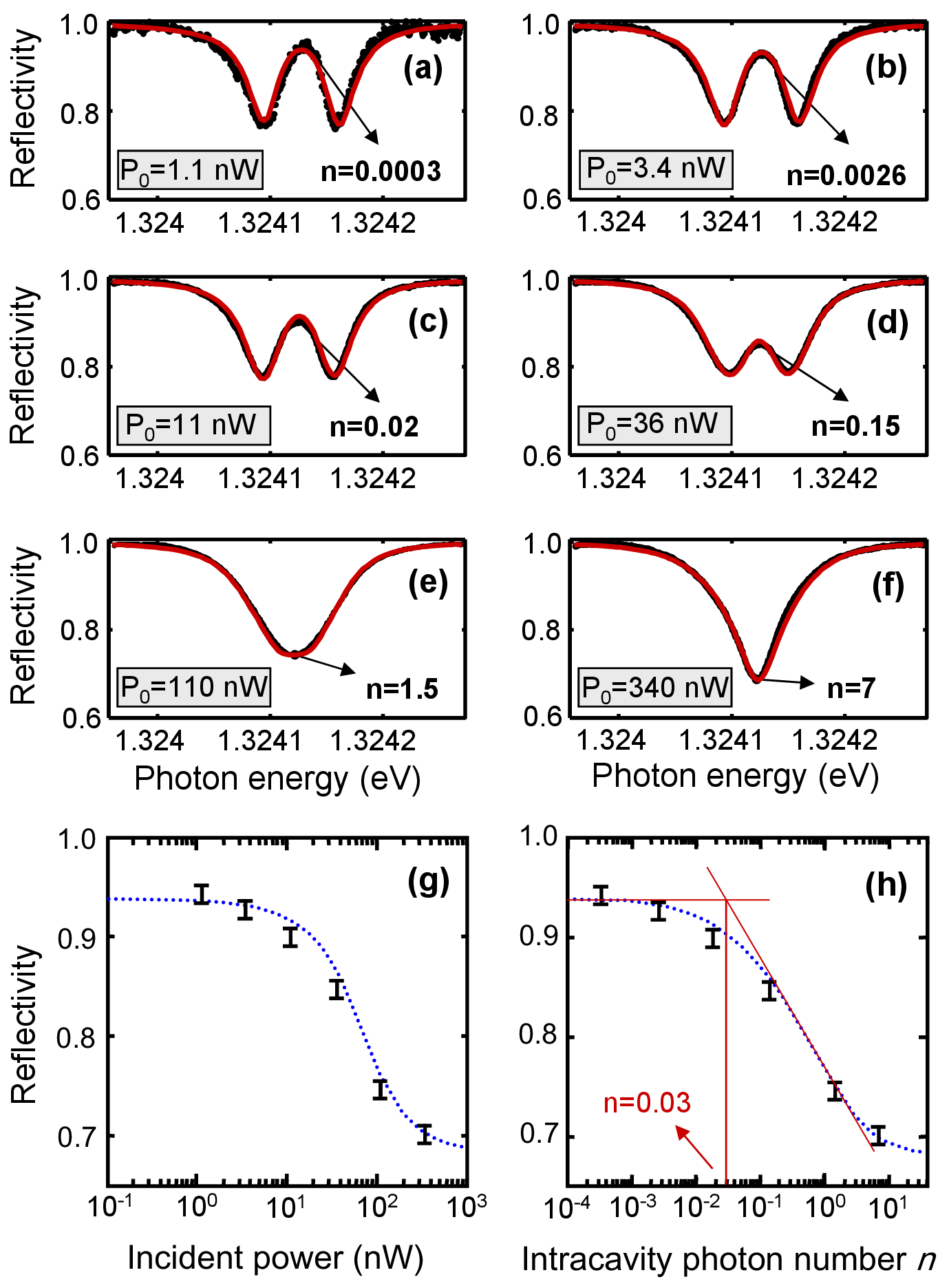}
\caption{\textbf{(a)} to \textbf{(f)} Reflectivity spectra at $T=35.9K$ for various incident powers $P_0$. Black dots: experimental points. Red solid lines: theoretical predictions. \textbf{(g)} (resp. \textbf{(h)}) Reflectivity at resonance as a function of $P_0$ (resp. $n$). Black dots with error bars: experimental points. Blue dotted lines: theoretical predictions. Red straight lines: guides to the eyes for the nonlinearity threshold. \label{Fig3}}
\end{figure}

Figs. \ref{Fig3}(a) to \ref{Fig3}(e) display six spectra measured at a fixed temperature $T=35.9K$ for increasing values of the CW incident power $P_0$. A continuous transition is observed between the low-power regime with two reflectivity dips and the high-power regime with a single Lorentzian dip \cite{Loo2010}. Numerical fits are obtained by comparing the measured reflected powers for the six spectra with the predictions of cavity QED, with a single set of device parameters. The extracted cavity damping rate is $\kappa = 45 \mathrm{\mu eV}$ corresponding to a fairly high quality factor $Q = \frac{\omega_C}{\kappa}=29000$. The QD-cavity coupling strength is $g = 33 \mathrm{\mu eV}$, and the figure of merit for the strong-coupling regime is $4g/\kappa = 2.9$. A QD dephasing rate $\gamma_\bot= 10 \mathrm{\mu eV}$ is also extracted, corresponding to a cooperativity parameter $C = \frac{g^2}{\kappa \gamma_{\bot}} =2.5$; this is twice higher than previous record values obtained in QD-micropillar systems \cite{Rakher2009,Loo2010,Young2011}. 

The fits presented in Figs. \ref{Fig3}(a) to \ref{Fig3}(e) are obtained for an input coupling efficiency $\eta_\mathrm{in}=0.95$, consistent with the previously reported values \cite{Loo2010,Arnold2012}, an output coupling efficiency  $\eta_\mathrm{out}=0.16$, and a critical photon number $n_c=0.035$. The output coupling efficiency corresponds to a mirror damping rate $\kappa_m=3.6 \mathrm{\mu eV}$ for each mirror and thus to a planar quality factor $Q_0=\frac{\omega_C}{2 \kappa_m}=190000$. As for the critical photon number $n_c=\frac{\gamma_{||} \gamma_{\bot}}{4g^2}$, it corresponds to $\frac{\gamma_{||}}{2}= 7.5 \mathrm{\mu eV}$ and thus $\gamma^*= 2.5 \mathrm{\mu eV}$. We point out that  $g$, $\kappa$ and $\gamma_{\bot}$ can be directly deduced from low-power measurements only \cite{Rakher2009,Loo2010,Young2011}, but that the precise determination of $\eta_\mathrm{in}$, $\eta_\mathrm{out}$ and $n_c$ requires a specific fitting procedure taking into account the complete set of experimental data \cite{SupplementalMaterials}.

Whereas the pure dephasing rate $\gamma^*$ is typical to QD-cavity experiments in a range of temperature around 35 K, the incoherent decay rate $\gamma_{||}$ corresponds to an unexpectedly short lifetime of 45 ps. This is an order of magnitude lower than the lifetime expected for spontaneous emission into free-space optical modes (around 700 $ps$). This high value of $\gamma_{||}$ cannot be accounted for by non-radiative processes \cite{Stobbe2010} since the quantum dot line under investigation has demonstrated high emission rates in photoluminescence measurements. However, due to the micropillar ellipticity and quite large size, other cavity resonances are within a few meV from the optical mode under investigation. The large value of $\gamma_{||}$ can thus be explained by the Purcell-accelerated, phonon-assisted emission of photons into these other cavity modes, as previously evidenced in photonic crystal cavities \cite{Hohenester2009,Hughes2011}. To improve the system cooperativity and critical photon number, this phonon-assisted emission should be decreased by designing a device operating at a lower temperature and/or having no optical modes spectrally close to the fundamental one.

With the knowledge of the coupling efficiencies, it becomes possible to extract the intracavity photon number directly from the measured reflectivity \cite{SupplementalMaterials}. The values of the mean intracavity photon number $n$, as measured from the reflectivity spectra at resonance, are indicated in Figs. \ref{Fig3}(a) to \ref{Fig3}(e): the nonlinear behaviour appears at an intracavity photon number much lower than unity. Fig. \ref{Fig3}(g) (resp. \ref{Fig3}(h)) presents the values of the measured reflectivities at resonance, as a function of the incident power $P_0$ (resp. intracavity photon number $n$), together with the numerical predictions obtained with the above-mentionned set of parameters.  We stress that there is a nonlinear relation between $P_0$ and $n$: a factor 3 increase in $P_0$ can lead to a factor 10 increase in the intracavity photon number $n$. A nonlinear feedback is indeed present, due to the fact that the intracavity photon number governs the reflection coefficient, which in turn governs the mode ability to accept new intracavity photons. The amplitude of this effect is directly related to the high value of the cooperativity $C$, which implies a high contrast between the low-power and high-power reflectivities.

With these data it is possible to define the threshold of the optical nonlinearity, by tracing the tangent to the reflectivity curve at its inflexion point (where the nonlinearity is the most efficient) and taking its intersection with the low-power plateau. This nonlinearity threshold is obtained at $n\approx0.03$, close to the value of the critical photon number $n_c$. It means that, in principle, nonlinear all-optical switching can be performed with single-photon pulses: this is the final goal of several recent works using QD-cavity systems \cite{Englund2012,Sridharan2012,Volz2012}. However, even if the intracavity photon number $n$ is the relevant parameter governing the photon-photon interaction inside the cavity, the important quantity for practical applications is the number of incident photons per pulse in front of the focusing lens, $N$. The relation between $N$ and $n$ depends in a non-trivial way on the characteristics of the device, experimental setup, and pulse optical properties. 

\begin{figure}[ht!]
\centering \includegraphics[width=0.49\textwidth]{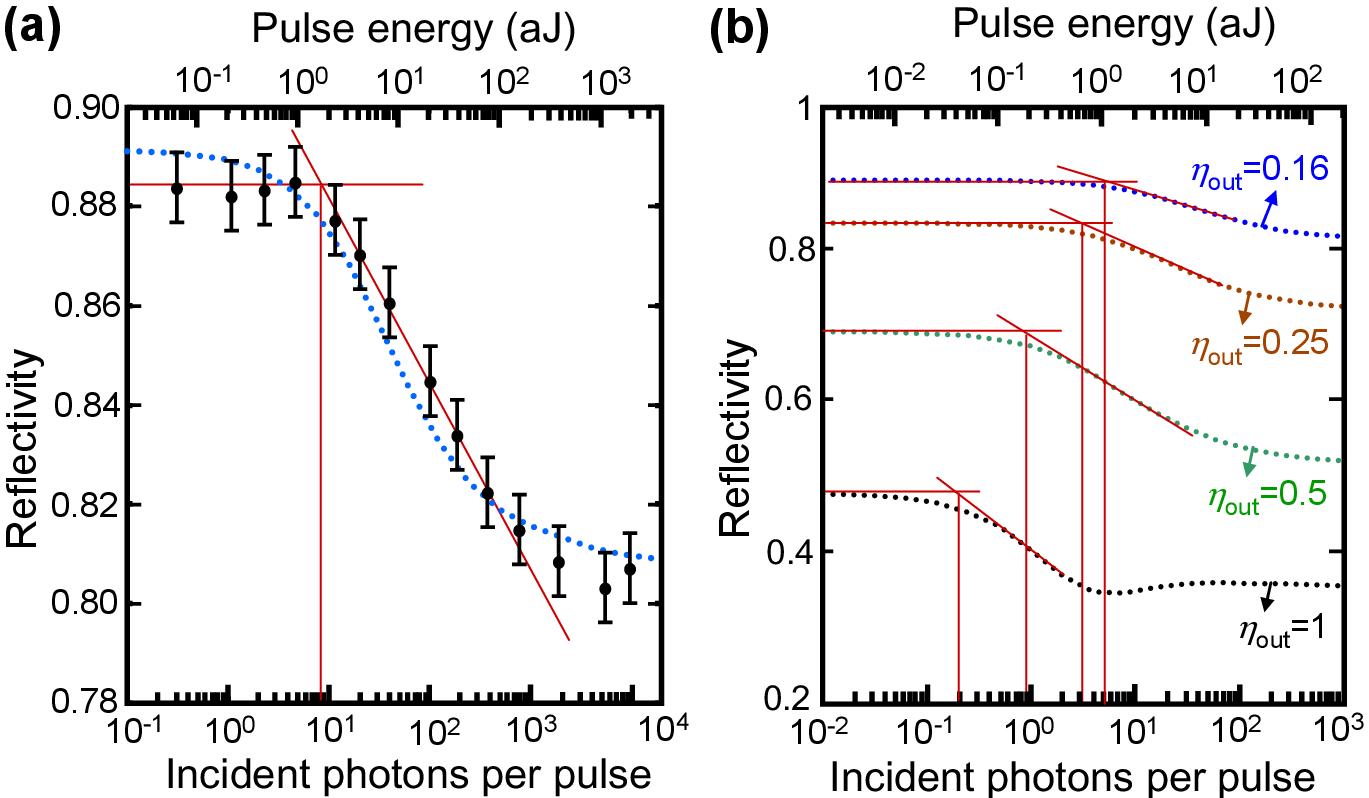}
\caption{\textbf{(a)} Reflectivity measurement under pulsed excitation as a function of the number of incident photons per pulse $N$. Black dots with error bars : experimental points. Blue dotted line : theoretical prediction. Red traight lines : guides to the eyes for the nonlinearity threshold. \textbf{(b)} Calculated reflectivity as a function of $N$ for various output coupling effiencies $\eta_\mathrm{out}$. Dotted lines: theoretical predictions. Red straight lines: guides to the eyes for the nonlinearity thresholds. \label{Fig4}}.
\end{figure}

To determine precisely the nonlinearity threshold in terms of $N$, reflectivity measurements were implemented under pulsed excitation while controlling the number of photons per pulse incident on the cavity. We used optical pulses with a $80$~MHz repetition rate and a pulse width $\Delta t=34$~ps, of the order of the photon lifetime in the cavity. The photon energy $\omega$ is tuned at resonance with both $\omega_c$ and $\omega_{QD}$, and the optical alignment and thus the input-coupling efficiency are left unchanged. The measured reflectivity is plotted in Fig.~\ref{Fig4}(a) as a function of $N$, together with the prediction of cavity QED using the above-mentionned device parameters. Following the same procedure as in Fig.~\ref{Fig3}(h) a nonlinear threshold of only $8$ incident photons per pulse is obtained, which is more than one order of magnitude lower than the recent values reported in quantum dot-photonic crystal systems \cite{Englund2012,Sridharan2012,Volz2012}. Here this ultralow nonlinearity threshold is obtained thanks to a near-unity input coupling efficiency. The fact that this threshold is still very different from the critical photon number $n_c=0.035$ illustrates how crucial it is to distinguish between $n$ and $N$ when working on the development of single-photon nonlinear devices.

In the perspective of further decreasing the nonlinearity threshold, we present in Fig.~\ref{Fig4}(b) numerical simulations showing that the output coupling efficiency $\eta_\mathrm{out}=0.16$ is the main limiting parameter. In this figure the expected reflectivity is plotted as a function of $N$ with all device and pulse parameters being kept constant, except the ratio $\frac{\kappa_m}{\kappa_s}$ and thus the output coupling efficiency $\eta_\mathrm{out}$. We find that the nonlinearity threshold is strongly influenced by this parameter, to the point that a factor six increase in $\kappa_m$ - and thus $\eta_\mathrm{out}$ - decreases the expected threshold by a factor 30. This non-linear dependence is related to the fact that $\kappa_m$ plays a role both in the injection of photons inside the cavity and in the collection of photons from the cavity mode. We stress that at near-unity output coupling efficiency a nonlinearity threshold lower than $1$ can be obtained, so that for $N=1$ incident photon per pulse the system will be precisely in the region of highest nonlinearity.

To conclude, we have realized a deterministically-coupled QD-micropillar system which behaves as a highly nonlinear medium when excited by few-photon optical pulses. Absolute reflectivity measurements have allowed a complete study of the giant optical nonlinearity and of the physical properties of the system. Thanks to the low critical photon number $n_c \approx 0.035$, and to the near-unity input coupling efficiency $\eta_\mathrm{in}\approx 0.95$, we obtained a record nonlinearity threshold of only 8 incident photons per pulse, measured in front of the focusing lens. This value, which constitutes the meaningful figure of merit, is the lowest ever reported. Several roads can be envisionned to improve the output coupling efficiency, and thus the nonlinearity threshold, including increasing the pillar diameter and/or designing adiabatic cavities for which the quality factor is more robust to micropillar sidewall imperfections \cite{Lermer2012}. The next crucial step, at the interface between nonlinear optics and quantum information, would be to provide a nonlinear interaction between two coincident single-photon pulses. This fundamental goal can be attained in realistic devices, through the development of strongly-coupled QD-micropillar systems with a near-unity output-coupling efficiency.

This work was partially supported by the French ANR JCJC MIND, ANR P3N CAFE, the ERC starting grant 277885 QD-CQED and the CHISTERA project SSQN.


%

\end{document}